\documentclass[reprint,secnumarabic,amssymb,superscriptaddress,aps, prl, twocolumn]{revtex4-2} 
\usepackage{amsmath,amssymb,bm}
\usepackage{graphicx}
\usepackage{color}
\usepackage[colorlinks=true,allcolors=blue]{hyperref}
\usepackage{xr}

\usepackage{dcolumn}
\usepackage[capitalize]{cleveref}
\usepackage{bm}

\newcommand{\beq}{\begin{eqnarray}}
\newcommand{\eeq}{\end{eqnarray}}

\def \beq{\begin{eqnarray}}
\def \eeq{\end{eqnarray}}

\begin{document}
\title{Enhanced stability and chaotic condensates in multi-species non-reciprocal mixtures}
\author{Laya Parkavousi}
\email{equal contributions}
\affiliation{Max Planck Institute for Dynamics and Self-Organization (MPI-DS), D-37077 Göttingen, Germany}
\author{Navdeep Rana}
\email{equal contributions}
\affiliation{Max Planck Institute for Dynamics and Self-Organization (MPI-DS), D-37077 Göttingen, Germany}
\author{Ramin Golestanian}
\email{ramin.golestanian@ds.mpg.de}
\affiliation{Max Planck Institute for Dynamics and Self-Organization (MPI-DS), D-37077 Göttingen, Germany}
\affiliation{Rudolf Peierls Centre for Theoretical Physics, University of Oxford, Oxford OX1 3PU, United Kingdom}
\author{Suropriya Saha}
\email{suropriya.saha@ds.mpg.de}
\affiliation{Max Planck Institute for Dynamics and Self-Organization (MPI-DS), D-37077 Göttingen, Germany}

\date{\today}

\begin{abstract}
Random non-reciprocal interactions between a large number of conserved densities are shown to enhance the stability of the system towards pattern formation. 
The enhanced stability is an exact result when the number of species approaches infinity and is confirmed numerically by simulations of the multi-species non-reciprocal Cahn-Hilliard model. Furthermore, the diversity in dynamical patterns increases with increasing number of components and novel steady states such as pulsating or spatiotemporally chaotic condensates are observed. Our results may help to unravel the mechanisms by which living systems self-organise via metabolism.
\end{abstract}
\maketitle

The cell cytosol solves an organisational challenge of baffling complexity in segregating large numbers of bio-molecules into functional units~\cite{Zimmerman1991}. In recent years, it has emerged that spatial organisation in cells is assisted by the formation of condensates driven by Liquid-Liquid Phase Separation (LLPS) \cite{SAHA20161572,PappuBrangwynne2015,Niebel2019,Franzmann2018}, a concept dating back to the protocells described by Oparin~\cite{oparin2003origin}. While it is evident that the desired combination of the overall agility of the sub-cellular environment, despite its relatively high packing density of generically sticky bio-molecules \cite{Parry2014,Bellotto2022}, and the dynamic structure formation must be consequences of non-equilibrium activity of the living cell, the mechanisms behind these features are still largely unknown. 

The equilibrium phase behaviour in a crowded environment with many interacting species has been studied extensively using interactions derived from free energies \cite{JacobsNmberOfComponents_PhysRevLett.126.258101,Jacobs2017,jacobs2013predicting,Shrinivas2021,qiang2024scalingphasecountmulticomponent,Weber2019,AndreK_PhysRevLett.125.218003,Joanny_Grosberg_PhysRevE.92.032118}. However, the appropriate theoretical description of dynamics and structure formation in a complex mixture with chemical cycles and gradients~\cite{Testa2021Enzyme} should incorporate non-equilibrium activity \cite{Gompper2020,marchetti2013}. More specifically, recent developments in theories of active phase separation in systems with {\it{finite}} numbers of species have examined the consequences of the competition between thermodynamic forces and various forms of non-equilibrium activity including self-propulsion~\cite{cates2015,cates2018,cates2019}, chemical activity~\cite{FreyErwin,Matthew-PRL-2022}, catalysis-induced self-phoresis~\cite{Golestanian2019phoretic,ramin-PRL-12,Prathyusha2024,agudo-canalejo2019,OuazanReboul2023NJP}, and non-reciprocal interactions \cite{saha2020,you2020, Dinelli2023}. 

\begin{figure}[b]
    \centering
    \includegraphics[width=\linewidth]{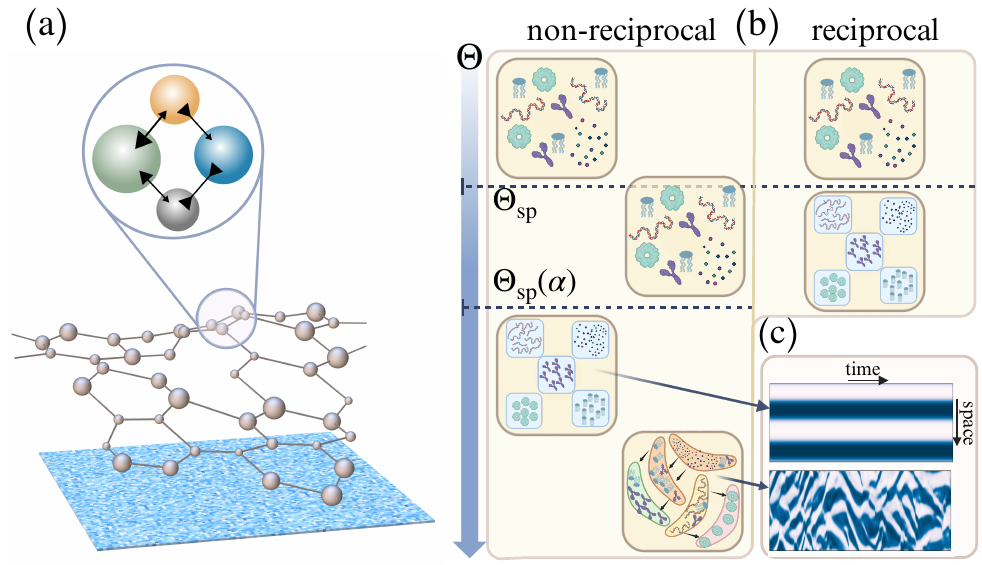}
    \caption{(a) Schematic representation of a minimal living cell, which we define as a collection of a large number of bio-molecular building blocks with random interactions. (b) Non-reciprocity is a reliable way to control LLPS, as it {\it{always}} lowers the phase separation temperature $\Theta$. (c) Activity stirs the mixture while simultaneously stabilising it, leading to chaotic dynamics as shown in the representative kymographs.}
    \label{fig1:Schematic}
\end{figure}

Here, we examine whether a cell, defined minimally as a collection of a large number of bio-molecular building blocks with random interactions as depicted in \cref{fig1:Schematic}, can generically harness non-reciprocity to regulate its internal spatiotemporal dynamics towards its function. We develop a theoretical framework to study phase separation in a system with a large number of components with broken action-reaction symmetry as the only source of non-equilibrium activity. Non-reciprocity arises naturally in chemically active systems \cite{soto2014,saha2019,agudo-canalejo2019} and the corresponding active mixtures \cite{agudo-canalejo2019,Tucci2024}, and stems from a frustration in microscopic interactions~\cite{Ramin_10.1051epn2024305}. It has been shown to lead to the emergence of chirality~\cite{fruchart2021} in polar systems~\cite{Dadhichi_PhysRevE.101.052601}, parity- and time-reversal symmetry breaking in number conserving densities~\cite{saha2020, you2020, ThieleFirst, saha2022, rana2023,brauns2023nonreciprocal,Pisegna2024}, formation of memory~\cite{loos2020}, multifarious self-organisation~\cite{Osat2023}, and odd response~\cite{Scheibner2020, ActiveSolidBartolo, Tan2022}.

We assume random interactions between components, taking inspiration from previous examples of the applications of random matrix theory (RMT) \cite{mehta2004random, Majumdar2014}, e.g. characterization of the spectra of heavy nuclei as pioneered by Wigner~\cite{Wigner1951} and May's study of the stability of ecological networks \cite{MAY1972}. Other applications include spin glass physics~\cite{mezard1987spin,Crisanti_Sompolinsky_PhysRevA.37.4865,Crisanti_Sompolinsky_PhysRevA.36.4922}, diverse ecosystems~\cite{Blumenthal2024} and the generalized Lotka-Volterra equations~\cite{Baron2022,Baron2023}.
We study the multi-species Non-Reciprocal Cahn-Hilliard (NRCH) model \cite{saha2020} to predict how non-reciprocity regulates phase behaviour using tools from the RMT and extensive simulations (see \cref{fig1:Schematic}), and find that non-reciprocity stabilises the homogeneous mixture and leads to a rich variety of patterns, from traveling bands to chaotic condensates.

\begin{figure}[t]
    \centering
    \includegraphics[width=\linewidth]{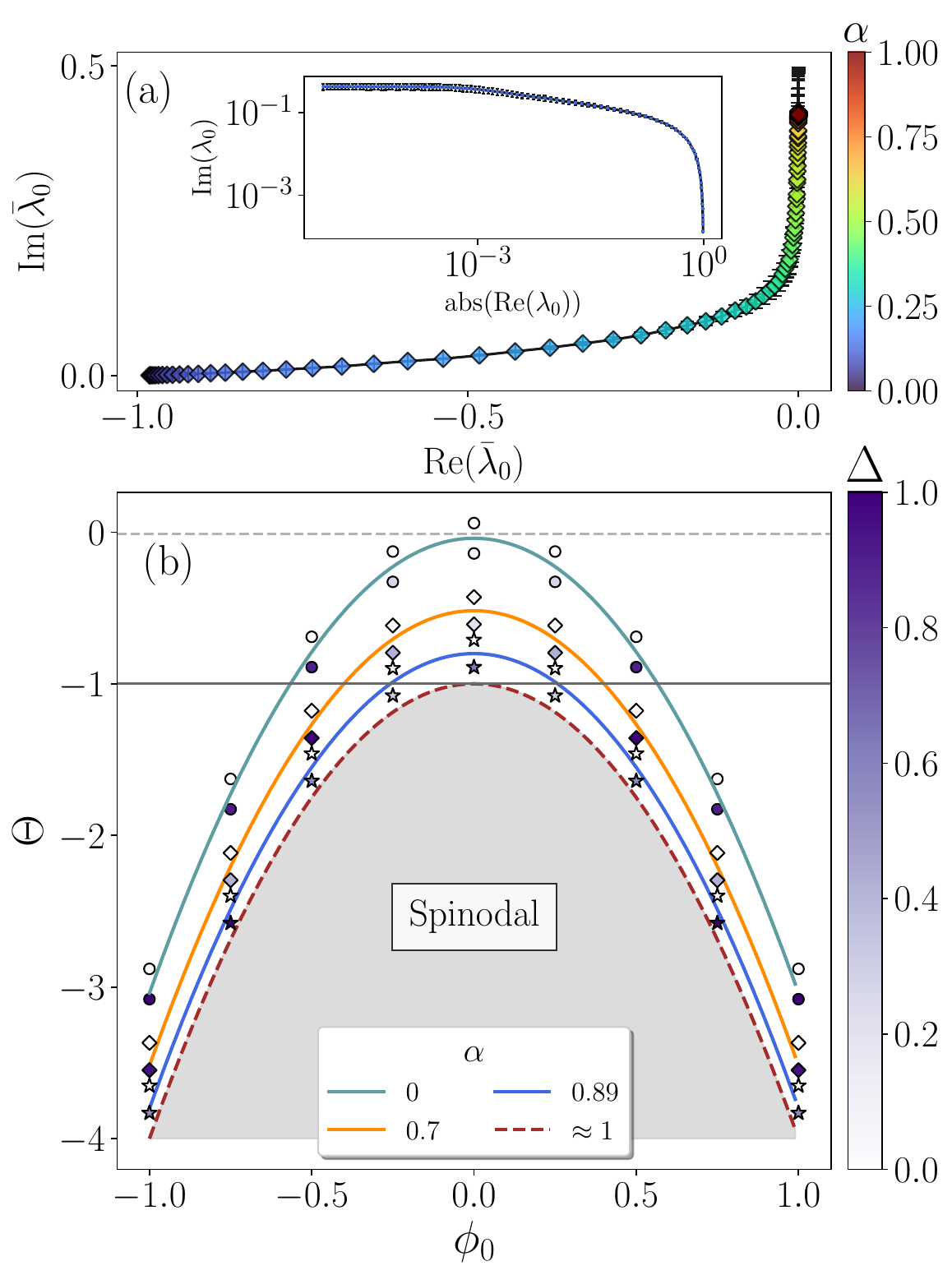}
    \caption{\label{fig:Spinodal} The eigenvalue of the interaction matrix $M_{ab}$ [see Eq.~\eqref{eq:M}] with the smallest real part, $\lambda_0$, determines the linear stability of the mixed state. (a) Scatter plots of the mean $\bar{\lambda}_0$ in the complex plane versus $\alpha$ for $N=10^3$ (error-bars show the variance). Inset: The same in log-log scale. (b) Solid lines are spinodal curves in the space spanned by temperature $\Theta$ and average composition $\phi_0$ at selected values of activity $\alpha$. The curves shift towards lower $\Theta$ at high $\alpha$ as $\mbox{Re}(\lambda_0)$ decreases showing that non-reciprocity stabilises the mixed state. Markers highlight the order parameter $\Delta$ [see Eq.~\eqref{eq:OrderParameter}], which vanishes outside the spinodal and changes discontinuously to a finite value inside it. The dashed line at maximum activity $\alpha \sim 1$ is the lower bound on the stable region. }
\end{figure}

\footnotetext[11]{See Supplemental Material available at xyz. It includes the Refs. \cite{canuto1988, cox2002, rana2023, Sprott2021, Edson2019}, a description of the numerical algorithm, the eigenvalue spectrum, the order parameter used to determine the numerical phase boundary, eigenvector corresponding to the fastest growing mode, and various measures used to classify the steady states.}

{\it{Theoretical framework.---}}We consider a mixture of $N$ individually conserved active species, which interact via non-reciprocal couplings \cite{Note11}. Number conservation implies that the scalar field $\phi_a({\bm r},t)$ associated with the number density of the $a$-th species evolves following the gradient of a non-equilibrium chemical potential $\mu_a$ as follows
\beq\label{eq:multiNRCH}
\partial_t \phi_a = \Gamma_a \nabla^2 \mu_a, \hskip3mm \mu_a = \frac{\delta F}{\delta \phi_a} + \alpha_{ab} \phi_b,
\eeq
where $a$ is assigned values from $1$ to $N$, $F$ is the free energy, $\Gamma_a$ is the mobility of species $a$, and summation over repeated indices is assumed. The system is maintained out of equilibrium by stipulating that in addition to interactions driven by a free energy that promotes macroscopic phase separation, the species interact via non-reciprocal interactions that are implemented 
via anti-symmetric coupling coefficients $\alpha_{ab} = -\alpha_{ba}$ 
\cite{saha2020}. To simplify a problem that is already complex at equilibrium (where $\alpha_{ab} = 0$)~\cite{Sollich2001}, here we consider only pair-wise inter-species interactions (ignoring multi-species interactions involving three or more species \cite{Zwicker_nonPairWise_PhysRevResearch.6.033002}), and choose the following form for the free energy $ F = \int_{\bm r} \left(\sum_a f_a+ f_{\rm FH}\right)$, where 
 \begin{equation}
f_a=\frac{\Theta_a}{2} \phi_a^2 + \frac{s_a^4}{4} \phi_a^4 + \frac{K_a}{2} | \nabla \phi_a|^2, \hskip3mm  f_{\rm FH}=\frac{\chi_{ab}}{2} \phi_a \phi_b,
 \label{eq:free-energy}
 \end{equation}
 where the elements of the symmetric matrix $\chi_{ab} = \chi_{ba}$ in the Flory-Huggins free energy density $f_{\rm FH}$ represent reciprocal interactions (we assume $\chi_{ab} =0$ for $a=b$). With appropriate re-scaling, we can set the couplings $s_a$ to unity. 
 The quadratic self-interaction term in Eq.~\eqref{eq:free-energy} is chosen to be identical for all species, i.e. $\Theta_a = 1+\Theta$, where $\Theta$ is an external control parameter similar to an effective temperature. Unequal self-interactions lead to competition and emergence of two or more strongly interacting groups of species~\cite{Filipe2023}. We note that $\Theta = -1$ is the critical temperature at which the non-interacting passive system ($\chi_{ab} = \alpha_{ab} = 0$) undergoes system wide spinodal decomposition for vanishing mean density. Finally, the coefficients of interfacial tension $K_a$ are chosen to be identical thereby avoiding a Turing instability~\cite{ThieleTuring}. $\sqrt{K}$ is used as the unit of length (and set to unity).

{\it{Linearised theory and the interaction matrix.---}}Using Eqs.~\eqref{eq:multiNRCH} and \eqref{eq:free-energy}, we find that the Fourier modes of density fluctuations $\delta \phi_a(q,t)$, defined as perturbations around the uniform value $\phi_0$ chosen to be identical for all species, evolve as follows
\beq\label{eq:LinDyn}
\dot{\delta \phi_{a}} = -q^2  \left[\left(1+\Theta  + 3 \phi_0^2\right) \delta_{ab} + M_{ab}\right] \delta \phi_{b}.
\eeq
Here, $M_{ab}  = \chi_{ab} + \alpha_{ab}$ is the interaction matrix and we have included terms up to quadratic order in $q$. Equations \eqref{eq:multiNRCH} and \eqref{eq:free-energy} still represent a complex system with an unknown stability manifold in a high dimensional parameter space spanned by the elements of the interaction matrix $M_{ab}$. We now consider ensembles of systems defined by a collection of interaction matrices $M_{ab}$ whose elements are drawn from Gaussian distributions with zero mean and variances $1$ and $\bar{\alpha}$ for the symmetric and anti-symmetric parts, respectively (thereby fixing the scale of reciprocal interactions to unity). The ensemble is represented as follows
\beq\label{eq:M}
M_{ab} = \frac{1}{\sqrt{N(1 + {\bar{\alpha}}^2)}} \left[ S_{ab} + {\bar{\alpha}} A_{ab} \right], 
\eeq 
for $a \neq b$ ($M_{ab}=0$ when $a=b$), where $S_{ab}$ and $A_{ab}$ are symmetric and anti-symmetric square matrices of size $N$, respectively, whose coefficients are drawn from a Gaussian distribution with zero mean and unit variance. Note that $M_{ab}$ is scaled such that the variance of its eigenvalues, $\lambda_i$, is bounded by unity. By redefining the non-reciprocity coupling constant as follows $\alpha(\bar{\alpha}) = \bar{\alpha}/\sqrt{N(1 + {\bar{\alpha}}^2))}$, we obtain a single control parameter $\alpha \in [0,1)$ that tunes the level of activity in the system. 

\begin{figure}[t]
    \centering
    \includegraphics[width=0.99\linewidth]{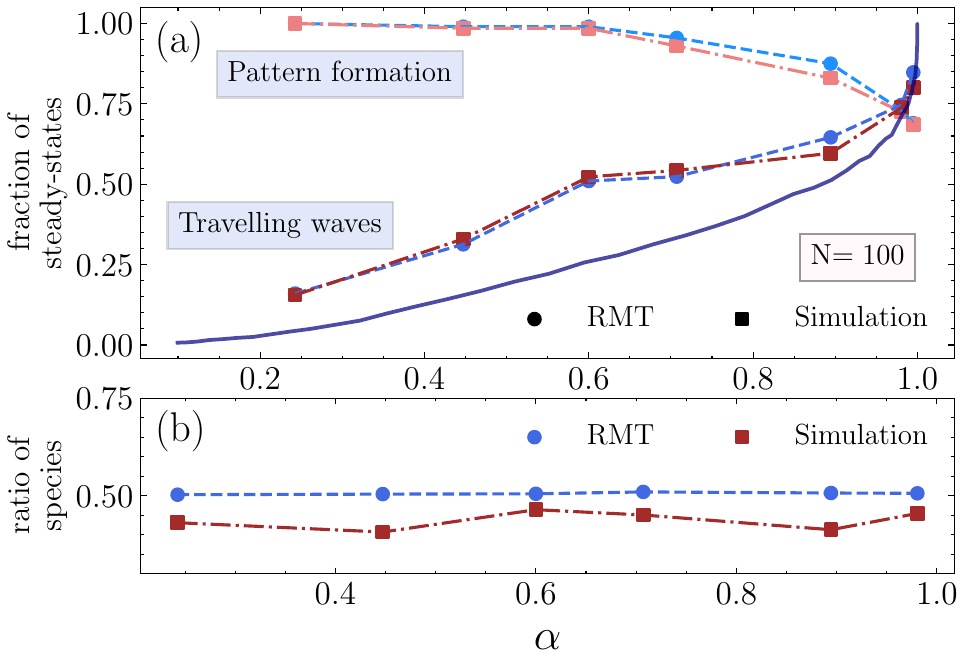}
    \caption{Statistics of steady states close to the spinodal. (a) Probability of finding patterns and waves in the steady state. Simulations results match well with the predictions of the linear theory for $\Theta(\alpha,0) <\approx \Theta_{\rm{sp}}(\alpha,0)$. The round markers and the blue line are results from an ensemble of size $300$ and $3\times10^5$ showing that the results converge very quickly with size of the ensemble. (b) The composition of the bulk phase separated final states as determined from simulations is compared with predictions from the eigenvector associated with $\lambda_0$.  }
    \label{fig:Statistics}
\end{figure}

The eigenvalue spectrum of $M_{ab}$ is given by the generalized semi-circle law~\cite{sommers1988}, the eigenvalues are distributed uniformly with a probability $1/(4 \pi \alpha)$ within an ellipse in the complex plane centred at the origin and with major and minor axes of sizes $2$ and $2\alpha^2$, respectively. 
For a symmetric random matrix ($\alpha=0$) the distribution reduces to the Wigner’s semi-circle law \cite{benaych2016lectures, wigner2,wigner1993characteristic, mehta2004random}. $\lambda_0$ denotes the most unstable eigenvalue of $M_{ab}$, i.e. the eigenvalue with the smallest real part. The distribution of $\lambda_0$ is bounded with a well-defined mean and variance that decreases as $1/\sqrt{N}$ with increasing $N$ \cite{Tracy1994}.

{\it{Threshold of the spinodal instability.---}}We define the ensemble-averaged value of the threshold, $\Theta_{\rm{sp}}(\alpha,\phi_0)$, as the value of $\Theta$ at which the largest eigenvalue of the dynamical matrix in Eq.~\eqref{eq:LinDyn} vanishes, thereby rendering the homogeneous state unstable, namely
\begin{eqnarray}
 \Theta_{\rm{sp}}(\alpha,\phi_0)  &=& - 3 {\phi_0}^2  -1 -  \mathrm{Re}[\bar{\lambda}_{0}(\alpha)].
 \label{eq:spinodal}
\end{eqnarray}
Equation \eqref{eq:spinodal} represents a family of parabolic curves that shift towards lower $\Theta$ upon increasing $\alpha$; see the solid curves in Fig.~\ref{fig:Spinodal}(b). Since the variance of $\Theta_{\rm{sp}}$ decays as $N^{-1/2}$ in the large $N$ limit, $\mbox{Re}(\lambda_0)$ converges to the ensemble-averaged values plotted in Fig.~\ref{fig:Spinodal}(a), implying that changing the exact values of $M_{ab}$ while complying with the definition in Eq.~\eqref{eq:M} does not appreciably alter the result. A comparison of the stability of the system at equilibrium with that at the highest value of activity at vanishing $\phi_0$ gives
\begin{eqnarray}
\Theta_{\rm{sp}}(0,0) = 0, \hskip5mm\Theta_{\rm{sp}}( 1,0) = -1.
\label{eq:stability}
\end{eqnarray}
Attractive (repulsive) interactions between species lead to pattern formation at higher (lower) $\Theta_{\rm{sp}}$. Recalling that $\Theta_{\rm{sp}}$ for the non-interacting system is $-1$, we first note that random reciprocal interactions shifts the spinodal to $0$, which can be interpreted as being influenced by an domination of attractive interactions. Random non-reciprocal interactions reduce the strength of the effective attractions, thereby lowering the threshold to the level of a non-interacting system at the maximum value of $\alpha$. This means that activity in the form of non-reciprocal interactions stabilises the system. 

The predictions from the linear theory are verified numerically by a simulation of the multi-species NRCH model in Eq.~\eqref{eq:multiNRCH} and \eqref{eq:free-energy} by tuning tuning the parameters $\phi_0$ and $\Theta$ to values just below and above the spinodal for $100$ realisations of $M_{ab}$ [see \cref{fig:Spinodal}(b)]. Mean squared deviation from the uniform composition $\Delta$ defined as 
\beq\label{eq:OrderParameter}
\Delta =  \int \mbox{d}x \left( \frac{1}{N}\sum_a\phi_a(x,t)- \phi_0 \right)^2,
\eeq
is used as a metric to identify the spinodal boundary, as it develops an abrupt jump; see \cref{fig:Spinodal}(b) and \cite{Note11}. 
 \begin{figure*}[t]
    \centering
    \includegraphics[width=0.95\linewidth]{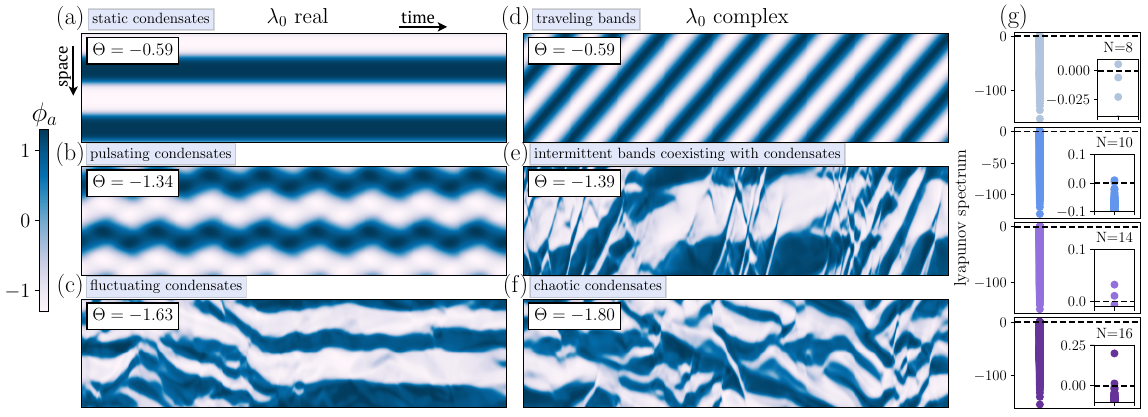}
    \caption{Pattern formation with decreasing $\Theta$. Panels (a-c) and (d-f) illustrate the dynamics in the steady state for two realisations of $M_{ab}$ ($\alpha = 0.89$), with real and complex $\lambda_0$, respectively. (a) For $\Theta<\Theta_{\rm sp}$ the system with vanishing $\phi_0$ shows bulk phase separation where roughly half of the species are enhanced in one of the phases. As $\Theta$ is lowered, the condensates pulsate with a finite number of distinct frequencies in the steady state as seen in panel (b). Upon lowering $\Theta$ further the oscillations become incoherent and the system starts to exhibit chaotic behaviour. In panel (e) waves (in which all species travel with fixed phase differences) give way to more complex chaotic dynamics. We find a zoo of complex dynamics that can be broadly categorised as intermittent bands and condensates in panel (e) and simply chaotic condensates in (f). (g) At sufficiently low $\Theta$ we only find chaotic condensates as characterised by the Lyapunov spectrum with at least one positive index. 
    }
    \label{fig:Patterns}
\end{figure*}

{\it{Steady states close to the spinodal.---}}For $\Theta \approx {\Theta}_{\rm sp}$ [see Eq.~\eqref{eq:spinodal}] and in the unstable region either one or a pair of the eigenmodes of the linearised system Eq.~\eqref{eq:LinDyn} grows exponentially depending on whether $\lambda_0$ is real or complex. This implies that close to the spinodal the multicomponent system with all its complexity can be compared to a binary system that phase separates either into bulk phases or travelling waves. From the distribution of $\lambda_0$ alone, we can decipher whether or not the unstable inhomogeneous state evolves to a steady state that breaks parity and time reversal symmetry. 

At selected values of $\alpha$, we consider ensembles of the matrix $M_{ab}$ with $n$ realisations and the corresponding set of $\lambda_0$. We fix $\Theta(\alpha) = 1.05\, \Theta_{\rm{sp}}(\alpha)$ and solve Eqs.~\eqref{eq:multiNRCH} and \eqref{eq:free-energy} numerically in one dimension starting from random initial configurations of the fields allowing the configurations to evolve to their steady states. A total of $n_{\rm{patterns}}$ of the $n$ steady state configurations are patterned and $n_{\rm{waves}}=n-n_{\rm{patterns}}$ are travelling waves. We compare the numerical findings to predictions from the linear stability analysis and summarise them in Fig.~\ref{fig:Statistics}. $n_{\rm{patterns}}$ is compared with the total number of realisations with $1+\mbox{Re}(\lambda_0) + \Theta < 0$ in Fig.~\ref{fig:Statistics}(a). $n_{\rm{waves}}$ is compared with the number of realisations with complex $\lambda_0$ in Fig~\ref{fig:Statistics}(a).

In the linear regime given by Eq.~\eqref{eq:LinDyn}, density fluctuations grow exponentially with stoichiometry along the eigenvector $\epsilon_{0a}$ corresponding to $\lambda_0$. In a passive system, $\epsilon_{0a}$ is perpendicular to the vector $u_a \equiv (1,1,1,...,1)$ \cite{Sear_2003}. We find that the overlap between the vector $u_a$ and $\epsilon_{0a}$ vanishes at all values of $\alpha$, implying that half of the components of $\epsilon_{0a}$ predominantly negate the other half. Consequently, we expect that the bulk phase separation is of the type where 
the phases are mutually exclusive, a prediction verified in our simulations as seen in \cref{fig:Statistics}(b). 

{\it{Classification of the long time dynamics.---}}Lowering $\Theta$ below $\Theta_{\rm{sp}}$ reveals complex dynamical behaviour in our numerical simulations. To explore these phases systematically, we divide the ensemble of $M_{ab}$ into two classes, corresponding to real and complex $\lambda_0$. Within each class the behavioural patterns are qualitatively similar as illustrated in panels (a-f) of \cref{fig:Patterns}, with an example shown for each class.

For real $\lambda_0$, the systems exhibit bulk phase separation for $\Theta$ just below $\Theta_{\rm{sp}}$, as shown in Fig.~\ref{fig:Patterns}(a). A new dynamical steady state with features common to both travelling waves and condensates appears upon lowering $\Theta$, a state that we denote as pulsating condensate, shown in \cref{fig:Patterns}(b). The pulsating condensate has well defined spatial domains that oscillate in time around a mean position with a fixed frequency. Upon further decreasing $\Theta$, the condensates fluctuate with more than a single frequency, a state that we denote as fluctuating condensates, shown in \cref{fig:Patterns}(c). 
For complex $\lambda_0$, we find travelling bands for $\Theta$ close to $\Theta_{\rm{sp}}$, which break the parity and time reversal symmetry of the phase separated domains. Density waves of all the species have the same wavelength, the phase difference between the species stays constant, and the bands move either to the left or to the right with a constant velocity, as shown in Fig.~\ref{fig:Patterns}(d). Upon decreasing $\Theta$, we find intermittent travelling bands that give way to condensates for large stretches of time before reforming again, as seen in Fig.~\ref{fig:Patterns}(e). With decreasing $\Theta$ further, we find spatiotemporal chaos, as shown in Fig.~\ref{fig:Patterns}(f), where a wide range of frequencies are involved as apparent from the power spectrum (see Supplemental Material \cite{Note11}). 
Finally, in Fig.~\ref{fig:Patterns}(g) we show the full Lyapunov spectra calculated for the number of components in the legend, verifying the chaotic behaviour of the dominant modes. We find that the number of positive indices rises from 1 to 2 as $N$ is increased from $8$ to $16$, while the magnitude of the largest index simultaneously increases. 

{\it{Concluding remarks.---}}We have studied the multi-component NRCH model for a large number of species and illustrated the role of non-reciprocity in regulating complex phase behaviour. The self-averaging nature of the most unstable eigenvalues in our application of RMT has enabled us to make robust predictions. The enhancement of stability is a striking feature of the model, which can lend itself to controllable biological regulation of LLPS. 
Our results apply to a wide range of non-equilibrium systems, as non-reciprocity can emerge in suspensions with enzymatic activity~\cite{Agudo-Canalejo2018_1,Testa2021Enzyme}, mass conserving reaction diffusion systems~\cite{Min2Densities_Brauns2021, Wurthner2022}, phoretic active matter \cite{Golestanian2019phoretic,Tucci2024}, or quorum sensing mixtures \cite{BenoitYu2023, TailleurTactic_PRL}, and can be used to drive a system towards or away from pattern formation in a controllable way relevant to biological systems~\cite{Franzmann2018}. Our work can be extended by exploring phase-composition \cite{qiang2024scalingphasecountmulticomponent,Sollich_Polydisperse_PhysRevLett.80.1365, Filipe2023}, tuning interfacial tension~\cite{ThieleTuring, cates2018}, elasticity~\cite{Monica_Olvera_PhysRevE.109.054409,Zwicker_PhysRevX.14.021009}, and incorporating sparse interaction matrices~\cite{Semerjian2002}. Our results may shed some light on the mechanisms to create and maintain homeostatic conditions in metabolically active cells, as well as spontaneous formation of metabolic cycles at the origin of life \cite{OuazanReboul2023,VincentPRL,OuazanReboul2023NJP}.

\begin{acknowledgements}
This work has received support from the Max Planck School Matter to Life and the MaxSyn-Bio Consortium, which are jointly funded by the Federal Ministry of Education and Research (BMBF) of Germany, and the Max Planck Society.   
\end{acknowledgements}

\bibliography{biblio}
\end{document}


\title{Enhanced stability and chaotic condensates in multi-species non-reciprocal mixtures \\ {\em Supplemental Material}}
\author{Laya Parkavousi}
\affiliation{Max Planck Institute for Dynamics and Self-Organization (MPI-DS), D-37077 Göttingen, Germany}
\author{Navdeep Rana}
\affiliation{Max Planck Institute for Dynamics and Self-Organization (MPI-DS), D-37077 Göttingen, Germany}
\author{Ramin Golestanian}
\email{ramin.golestanian@ds.mpg.de}
\affiliation{Max Planck Institute for Dynamics and Self-Organization (MPI-DS), D-37077 Göttingen, Germany}
\affiliation{Rudolf Peierls Centre for Theoretical Physics, University of Oxford, Oxford OX1 3PU, United Kingdom}
\author{Suropriya Saha}
\email{suropriya.saha@ds.mpg.de}
\affiliation{Max Planck Institute for Dynamics and Self-Organization (MPI-DS), D-37077 Göttingen, Germany}

\date{\today}
\maketitle

\tableofcontents

\setcounter{equation}{0}
\setcounter{figure}{0}
\renewcommand{\theequation}{S\arabic{equation}}
\renewcommand{\thefigure}{S\arabic{figure}}

\section{Numerical methods}
We use a pseudo-spectral algorithm to numerically integrate the $N$-species equations of motions given in the main text \cite{canuto1988}. We discretize the one-dimensional domain of length $L$ using $N_x$ equispaced collocation points. For time marching, we use second-order Exponential Time Differencing (ETD2) scheme \cite{cox2002, rana2023} with a fixed timestep $\Delta t$. 

We initialize our simulations with a homogeneous state perturbed with random perturbations drawn from a uniform distribution with zero mean. For a list of the parameters used in the simulations, see Table \ref{tab:parameter-list}.
To compute the Lyapunov spectrum, we adopt the algorithm proposed in \cite{Edson2019} for the Kuramoto-Sivashinsky equation. We have also verified that the algorithm gives correct Lyapunov exponents for the Lorenz attractor \cite{Sprott2021}.

\vspace{1cm}

\begin{table}[b]
 \begin{center}
\begin{tabular}{ p{6cm} p{3.5cm} p{3.5cm} p{3.5cm}  }
 \hline\hline
  & $N$&$L$& $\Delta t$\\
 \hline \hline
 Parameters in figure 1   & 100    &256&   $1.25\times 10^{-3}$\\
  Parameters in figure 2&   100  & 256   &$2.5\times 10^{-3}$\\
   Parameters in figure 3 &25 & 512&  $2.5\times 10^{-2}$\\

 \hline \hline
\end{tabular}
\end{center}
\caption{Parameter list}
 \label{tab:parameter-list}
\end{table}

\newpage
\section{Eigenvalues of $M_{ab}$}
\begin{figure*}[h]
    \centering
    \includegraphics[width=0.9\textwidth]{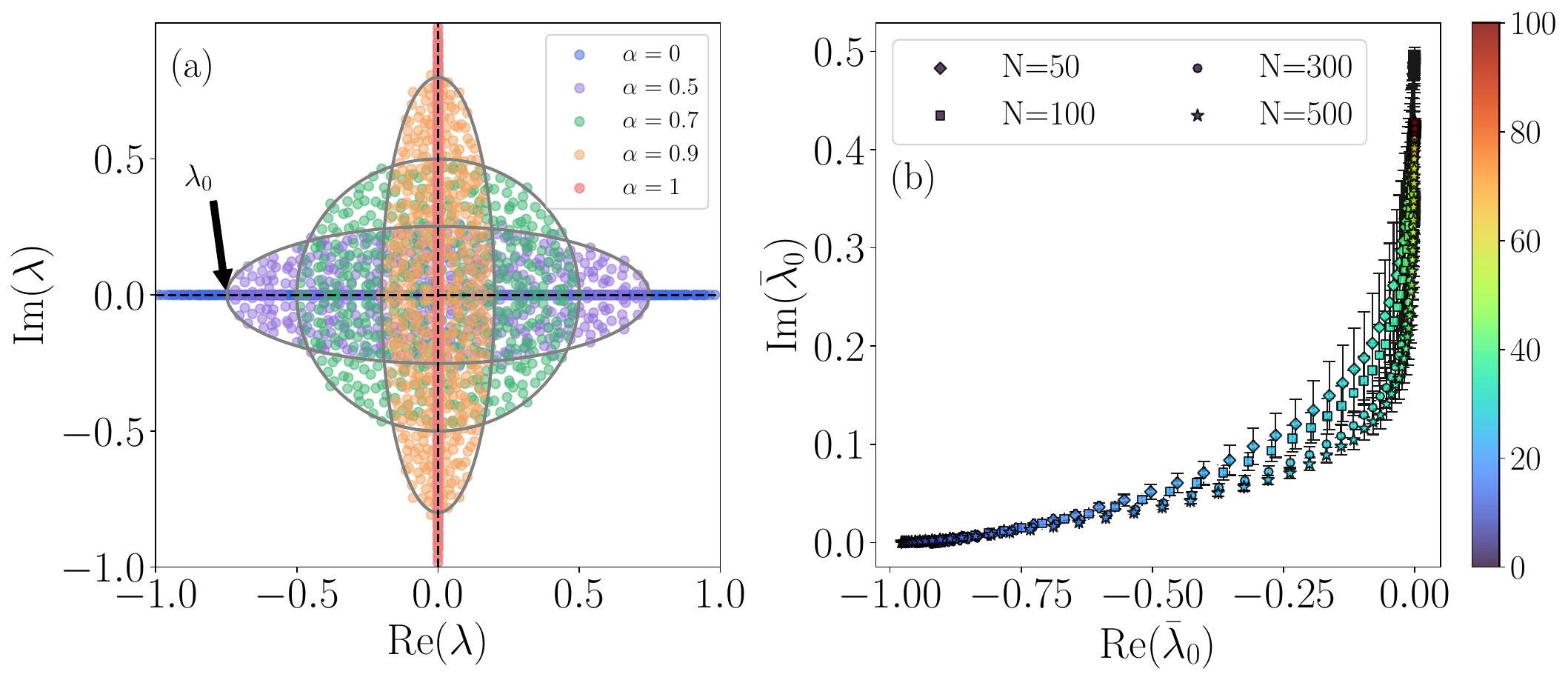}
    \caption{\label{figS1}(a) Numerical results for the distribution of eigenvalues for $N=100$ and ensemble size 500, for different values of the non-reciprocal parameter $\alpha$. (b) Variation of average value of real and imaginary components of $\lambda_0$, the eigenvalue with the smallest real part with $\alpha$, we compare random matrices with of different sizes $N$ for a fixed ensemble of size $3\times 10^4$. The color marks $\alpha$ and the error-bars show the variance. Note that the eigenvalues are re-scaled by $1/\sqrt{N(1 + \alpha^2)}$ .
    }
\end{figure*}

For a given ensemble size of $10^4$ and $N=100$, we have calculated the average smallest eigenvalue $\lambda_0$ and its real part for different values of $\alpha$. In \cref{figS1}, we plot the distribution of eigenvalues for various values of $\alpha$ in the complex plane.

\begin{figure*}[h]
    \centering
    \includegraphics[width=0.9\textwidth]{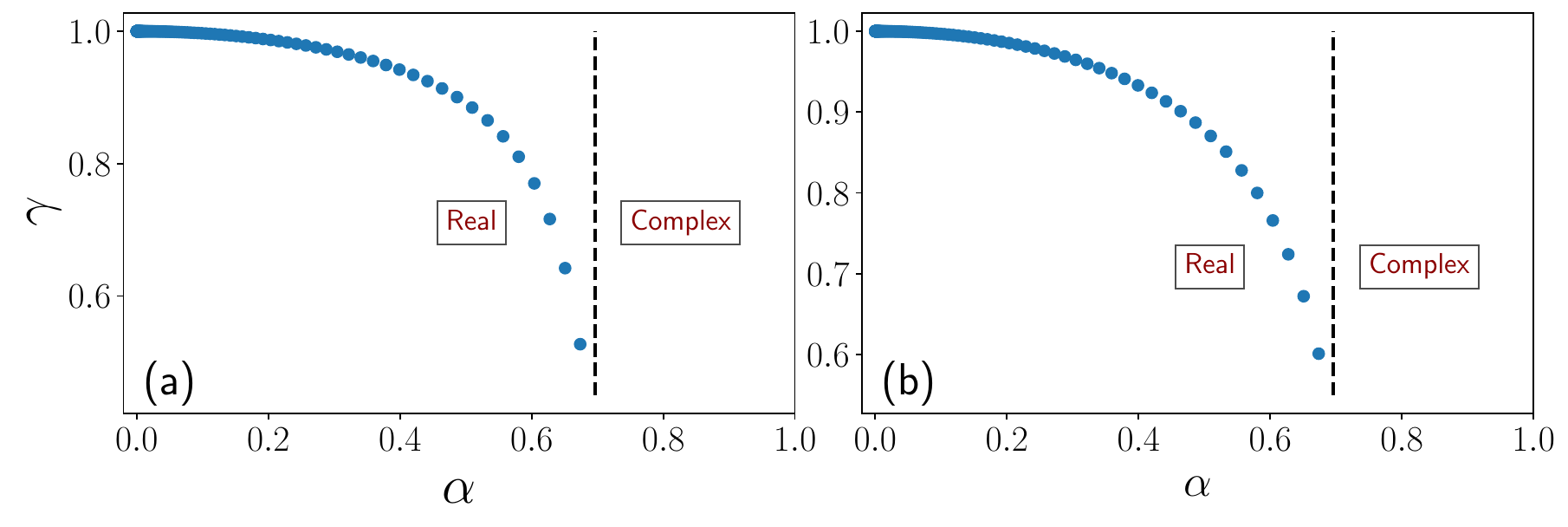}
    \caption{The overlap of the eigenvector $\epsilon_{0a}(\alpha)$ corresponding to $\lambda_0$ with $\epsilon_{0a}(0)$ as a function of $\alpha$. (a) and (b) show the overlap for the left and the right eigenvectors, respectively. }
    \label{eigenvector}
\end{figure*}

\newpage
\section{Determining the spinodal region}

\begin{figure*}[h]
    \centering
    \includegraphics[width=0.9\textwidth]{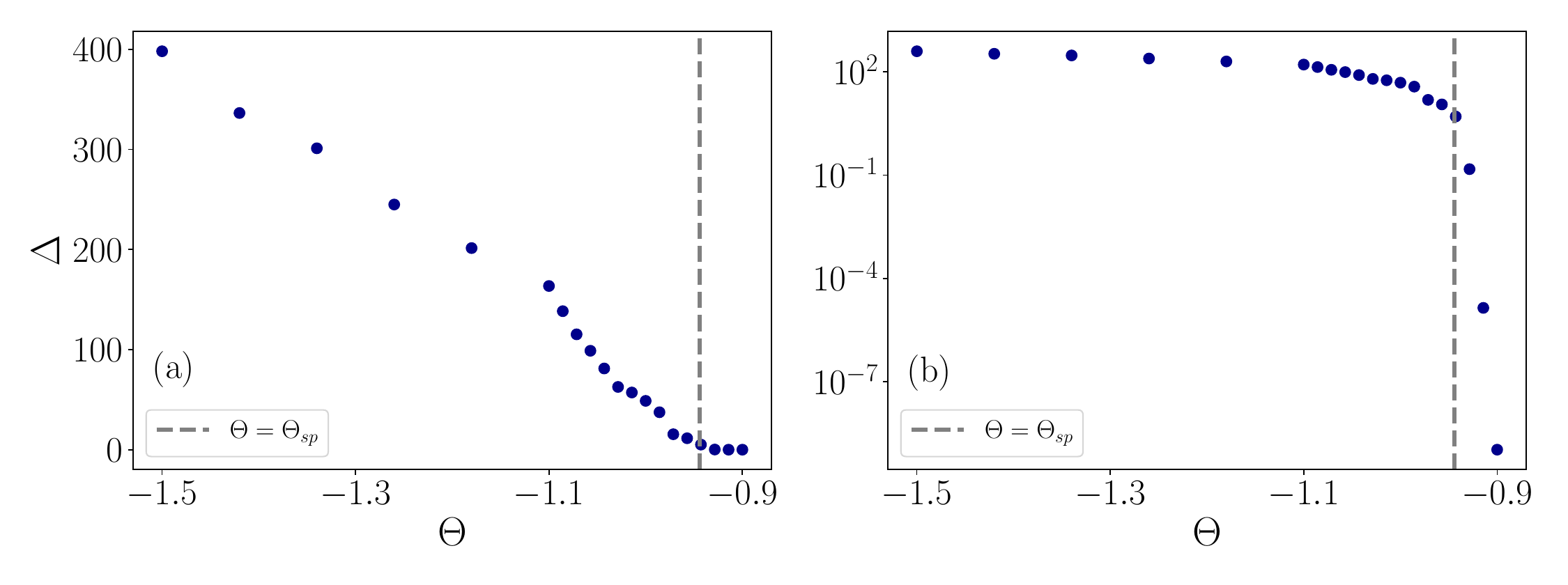}
    \caption{(a) The value of the measure $\Delta$ introduced in Fig. 2 in the main text for $\alpha = 0.89$. 
    (b) The same plot as panel (a) in a semi-log axis, to highlight the jump in $\Theta_{\rm sp}$. 
    All values of $\Delta$ are normalized by the largest value calculated. 
    }
    \label{figS3}
\end{figure*}

\newpage
\section{Characterisation of the dynamical steady states}

\begin{figure*}[h]
    \centering
    \includegraphics[width=\textwidth]{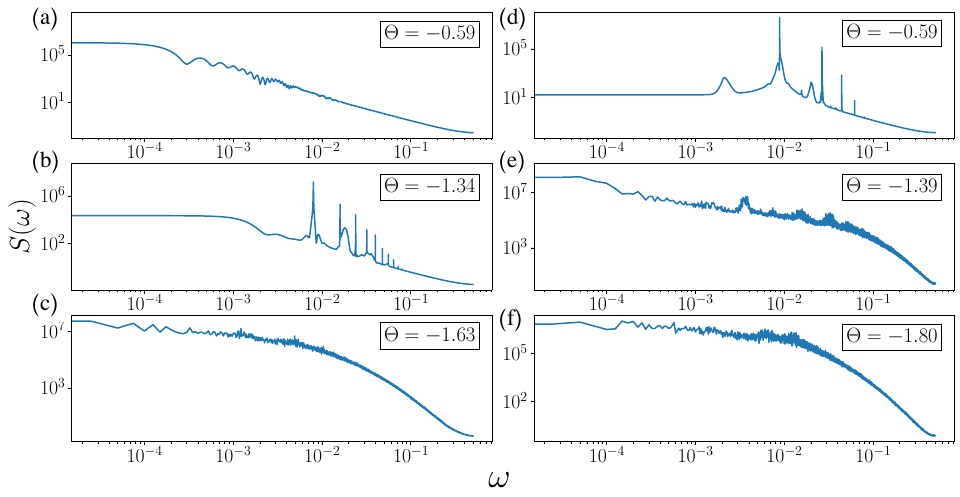}
    \caption{Power spectrum corresponding to the kymographs in Fig. 4 in the main text. For static condenstates (a), $S(\omega)$ is dominated by $\omega \to 0$ frequencies. For travelling bands (d), $S(\omega)$ shows dominant peaks at intermediate frequencies. Pulsating condensates (b) show similar peaks, but superimposed with a significant contribution from $\omega \to 0$. For intermittent, fluctuating and chaotic condensates (c,e,f), we find a broad range of active frequencies, with an increasing contribution from intermediate frequencies as we decrease $\Theta$.}
    \label{figS4}
\end{figure*}

We use the static structure factor, which we define as
\begin{equation}
S(\boldsymbol{q})=\frac{1}{T} \int_0^T \mathrm{~d} t \; {\phi}_i(\boldsymbol{q}, t) {\phi}_i(-\boldsymbol{q}, t) 
\end{equation}
and the power spectrum, defined as
\begin{equation}
S(\omega)=\frac{1}{A} \int \mathrm{d}^2 \boldsymbol{r} \; \phi_i(\boldsymbol{r},-\omega) \phi_i(\boldsymbol{r}, \omega) ,
\end{equation}
to characterise the dynamical steady states. See Fig. \ref{figS4} for the power spectrum corresponding to Fig. 4 in the main text.

\newpage
\begin{figure*}[h]
    \centering
    \includegraphics[width=\textwidth]{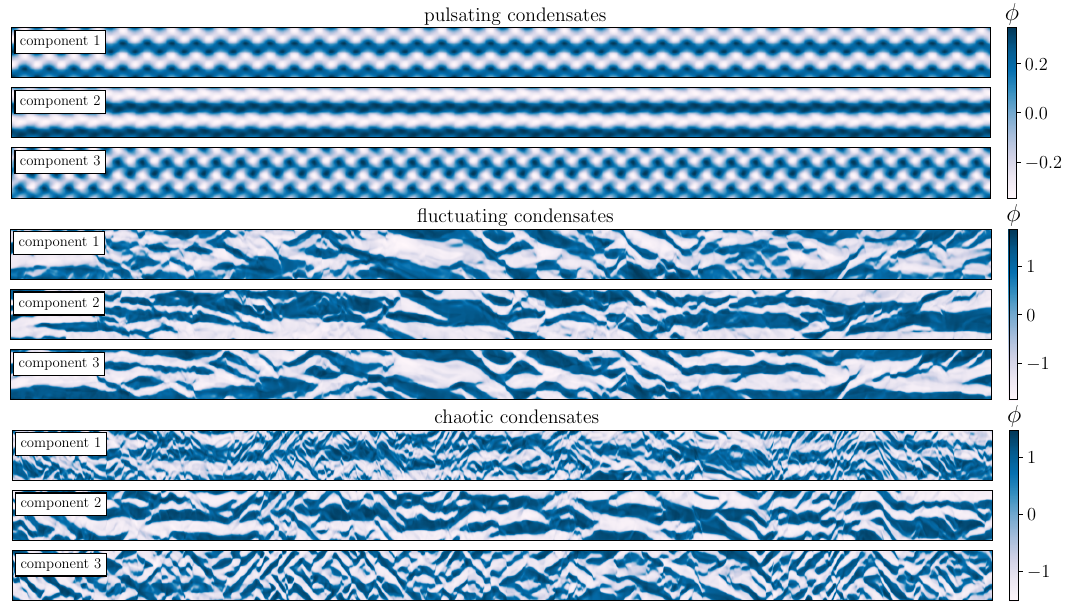}
    \caption{Kymographs of the different concentration fields in the NRCH model, for different categories of the observed condensates. For each condensate the kymographs for three of the components in the system are shown.}
    \label{figS5}
\end{figure*}

\bibliography{biblio}